\documentclass[conference,twoside,letterpaper]{IEEEtran}
\IEEEoverridecommandlockouts
\ifCLASSINFOpdf
  \usepackage[pdftex]{graphicx}
\else
\fi

\usepackage[cmex10]{amsmath}
\usepackage{cite}
\usepackage{mathtools}
\usepackage{flushend}

\begin{document}
%
% paper title
% can use linebreaks \\ within to get better formatting as desired
\title{Cancellation of Power Amplifier Induced Nonlinear Self-Interference in Full-Duplex Transceivers}
%
%
% author names and IEEE memberships
% note positions of commas and nonbreaking spaces ( ~ ) LaTeX will not break
% a structure at a ~ so this keeps an author's name from being broken across
% two lines.
% use \thanks{} to gain access to the first footnote area
% a separate \thanks must be used for each paragraph as LaTeX2e's \thanks
% was not built to handle multiple paragraphs
%

\author{\IEEEauthorblockN{Lauri~Anttila,
Dani~Korpi,
Ville~Syrj\"{a}l\"{a},
and~Mikko~Valkama}
\\
\IEEEauthorblockA{Department of Electronics and Communications Engineering, Tampere University of Technology, Finland\\ e-mail: lauri.anttila@tut.fi, dani.korpi@tut.fi, ville.syrjala@tut.fi, mikko.e.valkama@tut.fi}
\thanks{The research work leading to these results was funded by the Academy of Finland (under the projects \#259915, \#258364 "In-band Full-Duplex MIMO Transmission: A Breakthrough to High-Speed Low-Latency Mobile Networks"), the Finnish Funding Agency for Technology and Innovation (Tekes, under the project "Full-Duplex Cognitive Radio"), the Linz Center of Mechatronics (LCM) in the framework of the Austrian COMET-K2 programme, and Emil Aaltonen Foundation.}}% 
\maketitle

\begin{abstract}
%\boldmath
Recently, full-duplex (FD) communications with simultaneous transmission and reception on the same channel has been proposed. The FD receiver, however, suffers from inevitable self-interference (SI) from the much more powerful transmit signal. Analogue radio-frequency (RF) and baseband, as well as digital baseband, cancellation techniques have been proposed for suppressing the SI, but so far most of the studies have failed to take into account the inherent nonlinearities of the transmitter and receiver front-ends. To fill this gap, this article proposes a novel digital nonlinear interference cancellation technique to mitigate the power amplifier (PA) induced nonlinear SI in a FD transceiver. The technique is based on modeling the nonlinear SI channel, which is comprised of the nonlinear PA, the linear multipath SI channel, and the RF SI canceller, with a parallel Hammerstein nonlinearity. Stemming from the modeling, and appropriate parameter estimation, the known transmit data is then processed with the developed nonlinear parallel Hammerstein structure and suppressed from the receiver path at digital baseband. The results illustrate that with a given IIP3 figure for the PA, the proposed technique enables higher transmit power to be used compared to existing linear SI cancellation methods. Alternatively, for a given maximum transmit power level, a lower-quality PA (i.e., lower IIP3) can be used.
\end{abstract}
% IEEEtran.cls defaults to using nonbold math in the Abstract.
% This preserves the distinction between vectors and scalars. However,
% if the journal you are submitting to favors bold math in the abstract,
% then you can use LaTeX's standard command \boldmath at the very start
% of the abstract to achieve this. Many IEEE journals frown on math
% in the abstract anyway.

% Note that keywords are not normally used for peerreview papers.
%\begin{IEEEkeywords}
%Digital cancellation, full-duplex radio, nonlinear distortion, power amplifier, self-interference.
%\end{IEEEkeywords}

% For peer review papers, you can put extra information on the cover
% page as needed:
% \ifCLASSOPTIONpeerreview
% \begin{center} \bfseries EDICS Category: 3-BBND \end{center}
% \fi
%
% For peerreview papers, this IEEEtran command inserts a page break and
% creates the second title. It will be ignored for other modes.
\IEEEpeerreviewmaketitle

\section{Introduction}

\IEEEPARstart{F}{ULL}-DUPLEX communications using the same carrier for simultaneous transmission and reception has long been considered impossible due to the high self-interference (SI) from the transmitter to the receiver path. Recently, several research groups have experimentally demonstrated that, by utilizing various radio-frequency and digital baseband SI cancellation techniques, FD communications may indeed be possible with transmit powers in the local area communications (e.g. WiFi) range ($\leq$~20~dBm) \cite{Knox12,Choi10,Jain11,Sahai11,Duarte10,Bharadia13,Everett13}.

In \cite{Korpi13}, it was found that the nonlinear self-interference due to power amplifier (PA) nonlinearity may become a bottleneck in FD communications with higher transmit powers. Even though the transmit-receive antenna isolation and many RF SI cancellation techniques attenuate the PA-induced intermodulation distortion, such nonlinear SI can still be higher than the weak received signal and prevent successful detection \cite{Korpi13}. So far, only one of the digital SI cancellation techniques in the literature has considered these nonlinear effects explicitly \cite{Bharadia13}.\footnote{In \cite{Bharadia13}, a similar model as the one proposed in this article, was used to model the cascade of the PA and the multipath SI channel. This work was unavailable at the time of submission of our original manuscript.} The purpose of this article is to study the effects of PA nonlinearity on received signal quality, in full duplex transceiver context, with various amounts of RF cancellation (passive+active) and linear digital SI cancellation, and to propose \emph{a novel nonlinear SI cancellation technique to mitigate such nonlinear effects}. The proposed technique can be utilized in digital baseband or digitally-assisted analog SI cancellation. We demonstrate the performance improvements with full waveform simulations, showing that with typical RF front-end and signal parameters, up to 10 dB higher transmit power can be used without degrading the receiver signal-to-interference-plus-noise ratio (SINR) by using the proposed nonlinear SI cancellation technique instead of traditional linear SI cancellation. Alternatively, assuming some maximum transmit power level, utilizing the proposed technique enables using a PA with a lower IIP3 figure.

The rest of this paper is organized as follows. In Section~\ref{sec:challenges}, the challenges presented by a nonlinear PA in the context of full-duplex transceivers are discussed. In Section~\ref{sec:modeling}, the proposed nonlinear digital SI cancellation algorithm is presented. Then, in Section~\ref{sec:results}, the performance of the proposed algorithm is evaluated with waveform simulations. Finally, the conclusions are drawn in Section~\ref{sec:conc}.

\section{PA Nonlinearity Challenge in FD Transceiver}
\label{sec:challenges}

\begin{figure*}[!t]
\centering
\includegraphics[width=\textwidth]{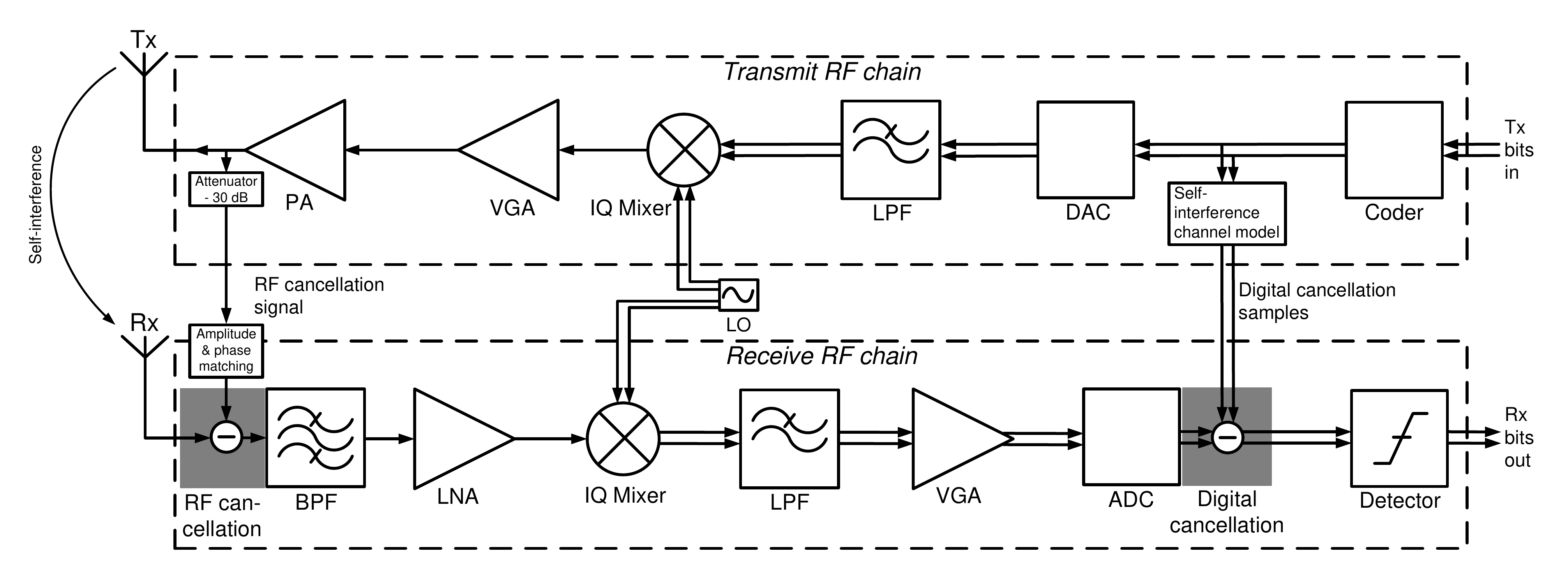}
\caption{A block diagram of the assumed full-duplex direct-conversion transceiver.}
\label{fig:block_diagram}
\end{figure*}

The assumed FD transceiver model is shown in Fig.~\ref{fig:block_diagram}. In this device, separate transmit and receive antennas provide natural isolation between the transmitter and receiver paths, and it is the basic FD transceiver model assumed in most of the existing literature \cite{Sahai11,Duarte10}. A notable exception is \cite{Knox12}, where a single-antenna device is assumed along with a circulator based isolation structure to provide passive cancellation of up to 40--45 dB over an 8 MHz band. However, the chosen FD model with separate transmit and receive antennas does not limit the applicability of the proposed techniques to this structure only, but the techniques can be applied in circulator based single-antenna FD devices as well.

In general, there can be four levels of SI cancellation in a FD device:

\begin{enumerate}
  \item Passive SI mitigation; the antenna isolation between transmit and receive antennas in a separate-antenna FD device, or the circulator isolation in a shared-antenna FD device
  \item Active spatial SI mitigation, i.e., beamforming in multi-antenna FD devices
  \item Active analog cancellation; SI cancellation in analog RF or baseband using the transmit signal as a reference
  \item Active digital cancellation; suppression of the residual SI with digital signal processing
\end{enumerate}

The state-of-the-art passive cancellation techniques, utilizing antenna design and placement techniques \cite{Choi10,Jain11,Sahai11,Duarte10}, or circulator design \cite{Knox12}, can yield up to 40--45 dB of cancellation. Active RF cancellation can give an extra 25--45 dB of SI cancellation, depending on the implementation \cite{Knox12,Choi10,Jain11,Sahai11,Duarte10,Bharadia13}, thus bringing the maximum achievable analog cancellation to the range of 65--90 dB. Up to this point the attenuation is the same for the linear and the nonlinear SI term. Then, digital baseband cancellation has been shown to bring another 20--30 dB of attenuation, at best \cite{Choi10,Jain11,Duarte10}. However, the digital cancellation techniques reported up to now have only considered linear SI, thus relying on fully linear signal processing. In the following, we first show with simple system power calculations that under realistic radio front-end and signal parameters from LTE user equipment, these cancellation levels are not enough to attenuate the power amplifier induced nonlinearities below the noise floor at higher transmit power levels. Motivated by this, we then propose novel nonlinear digital self-interference cancellation methods to suppress such distortion below the noise floor in Section~\ref{sec:modeling}.

Table~\ref{table:system_parameters} shows the baseline parameters of the transceiver used for the motivating system calculations, and also for the simulations. For the system calculations example, we assume 40 dB of passive cancellation, 30 dB of active analog cancellation, and a variable amount of linear digital cancellation to keep the linear SI term below the thermal noise floor at all power levels. These are somewhat optimistic assumptions, because for example the amount of achievable digital linear cancellation depends very much on the quality of the SI channel estimates, which in turn are affected greatly by the levels of nonlinear SI as well as thermal and quantization noise. The power level of the actual received signal of interest is assumed to be 5 dB above the sensitivity level in the following calculations. In addition, an automatic gain control algorithm is assumed to be tuning the gain of the VGA to match the dynamic range of the signal to the dynamic range of the analog-to-digital converter (ADC).

\begin{figure}[!t]
\centering
\includegraphics[width=\columnwidth]{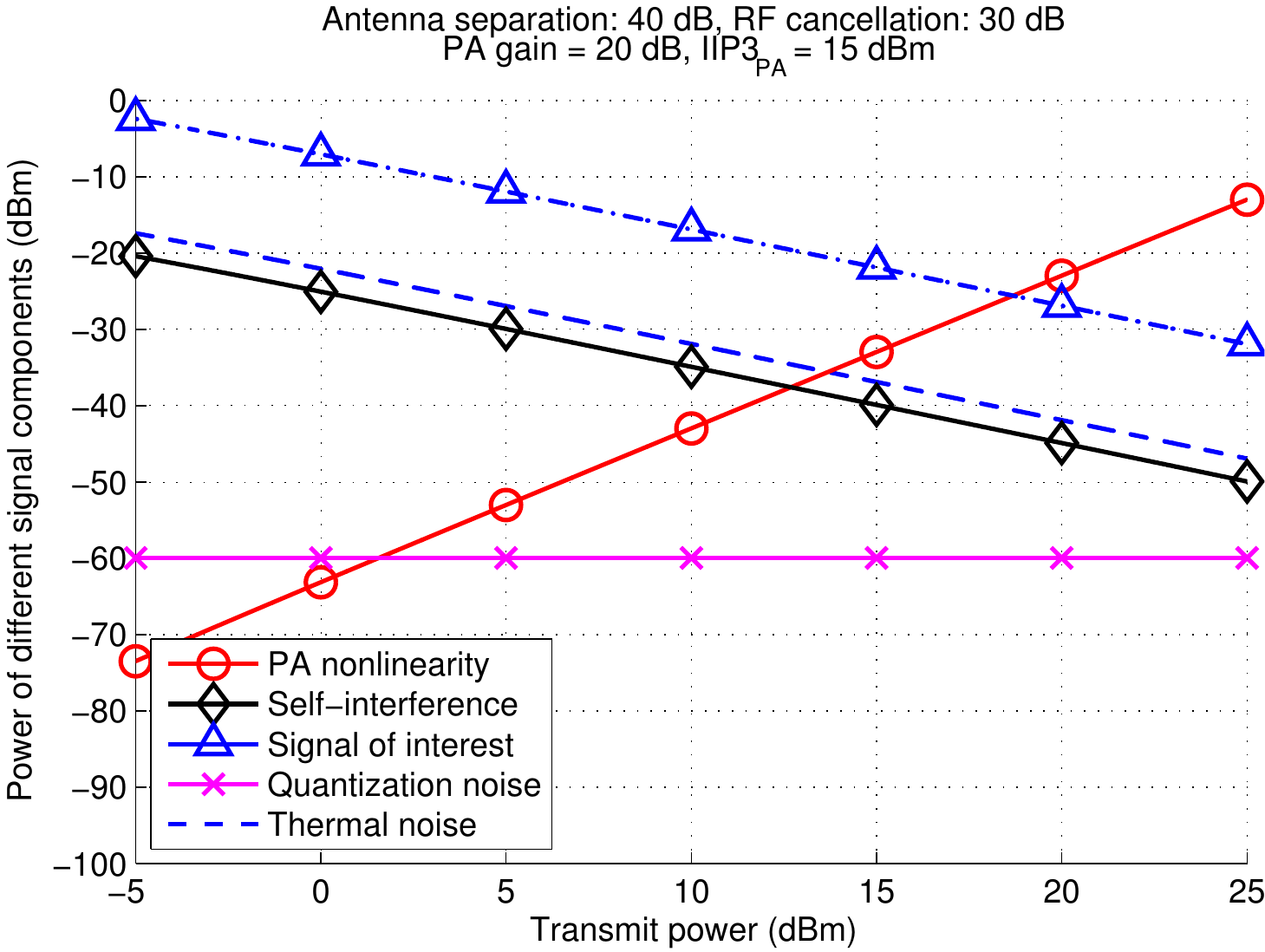}
\caption{An example plot of the power levels of the different signal components at receiver chain detector input with respect to transmit power.}
\label{fig:p_all_example}
\end{figure}

Fig.~\ref{fig:p_all_example} shows the powers of the different signal and interference terms at the detector input of the receiver. Despite the somewhat optimistic assumptions, the nonlinear SI, stemming from the PA nonlinearity, becomes the most powerful interference term already with transmit power of 14 dBm. With more realistic passive isolation and RF cancellation figures, the PA-induced nonlinear self-interference would be emphasized even further. Details of the analysis technique can be found from \cite{Korpi13}.

\begin{table}[!t]
\renewcommand{\arraystretch}{1.3}
\caption{The parameters of the full-duplex transceiver.}
\label{table:system_parameters}
\centering
\begin{tabular}{|c||c|}
\hline
\textbf{Parameter} & Value\\
\hline
SNR requirement & 10 dB \\
\hline
Bandwidth & 12.5 MHz\\
\hline
Receiver noise figure & 4.1 dB\\
\hline
Sensitivity & -88.9 dBm\\
\hline
Received signal power & -83.9 dBm\\
\hline
Antenna separation & 30/40/50 dB\\
\hline
RF cancellation & 30 dB\\
\hline
ADC bits & 12\\
\hline
ADC voltage range & 4.5 V\\
\hline
PAPR of the TX/RX waveform & 10 dB\\
\hline
PA gain & 20 dB\\
\hline
PA IIP3 & 15 dBm\\
\hline
PA 1 dB compression point & 24.5 dBm\\
\hline
\end{tabular}
\end{table}

%\begin{table}[!t]
%\renewcommand{\arraystretch}{1.3}
%\caption{Parameters for the relevant components of the transceiver.}
%\label{table:parameters}
%\centering
%\begin{tabular}{|c||c||c||c||c|}
%\hline
%\textbf{Component} & \textbf{Gain (dB)} & \textbf{IIP2 (dBm)} & \textbf{IIP3 (dBm)} & \textbf{NF (dB)}\\
%\hline
%PA (Tx)& 25 & - & 15 & 5\\
%\hline
%BPF (Rx)& 0 & - & - & 0\\
%\hline
%LNA (Rx) & 25& 43 & -15 & 4.1\\
%\hline
%Mixer (Rx)& 6 & 43 & 15 & 4\\
%\hline
%LPF (Rx) & 0 & - & - & 0 \\
%\hline
%VGA (Rx) & 0-69 & 43 & 10 & 4\\
%\hline\end{tabular}
%\end{table}

\section{Nonlinear Self-Interference Channel Modeling, Estimation and Cancellation}
\label{sec:modeling}

\subsection{Baseband nonlinear self-interference channel model}

We denote the original digital baseband transmit signal by $x_{n}$. For discrete-time baseband modeling of the nonlinear PA, we assume the widely-deployed parallel Hammerstein (PH) model, given as
\begin{align}
x_{n}^\mathit{PA} = \sum_{\substack{
   p = 1 \\
   p \text{ } \mathit{odd}
  }}^{P}
 \sum_{{k} = 0}^{{M}-1} f_{{p},{k}} \psi_{p}(x_{{n}-{k}}) \text{,}
\end{align}
where the basis functions are defined as $\psi_{p}(x_{n}) = |x_{n}|^{{p}-1} x_{n}$, $f_{p,n}$ are FIR filter impulse responses of the PH branches, $M$ denotes the memory length, and $P$ denotes the nonlinearity order of the PH model \cite{Ding04,Isaksson06,Anttila10}. The PH model has been shown to be a versatile tool for direct as well as inverse modeling of power amplifiers \cite{Ding04,Isaksson06,Anttila10,Ku03}.

The multipath SI channel between TX and RX antennas is modeled with an FIR filter $h_n$. RF cancellation aims to mitigate the main component of the multipath SI channel. To model possible frequency-dependency in the RF canceller path, due to for example delay mismatch or using an actual multitap RF canceller \cite{Bharadia13,Choi13}, we model the RF SI canceller response also with an FIR filter $a_n$ operating on $x_n^\mathit{PA}$. The received self-interference signal after the RF SI canceller can then be written as (with the nominal propagation delay removed)
\begin{align}
x_{n}^\mathit{SI} &= h_n \star x_n^\mathit{PA} - a_n \star x_n^\mathit{PA} \nonumber\\
&= \sum_{\substack{{p} = 1 \\ {p \text{ } odd}}}^{{P}}
\sum_{{k} = 0}^{{M}-1} \left( \left(h_k - a_k\right) \star f_{p,k}\right) \psi_{p}(x_{{n}-{k}}) \nonumber\\
&= \sum_{\substack{{p} = 1 \\ {p \text{ } odd}}}^{{P}}
\sum_{{k} = -M_1}^{{M}_2} \bar{f}_{{p},{k}} \psi_{p}(x_{{n}-{k}}) \label{eq:x_n_SI} \text{.}
\end{align}
Here, $\bar{f}_{{p},{k}}$ denote the \textit{effective model coefficients} of the overall nonlinear SI channel, and $M_1$ and $M_2$ are the non-causal and causal memory depths of the model, respectively. Therefore, the overall SI signal model, comprising of the nonlinear PA, the multipath SI channel, and the RF canceller, can also be expressed as a parallel Hammerstein model.% A block diagram of the canceller structure is presented in Fig.~\ref{fig:diag_canceller}. There, the different branches, corresponding to nonlinear terms of different order, can be seen, alongside with the coefficients $\bar{f}_{{p},{n}}$.

\begin{figure}[!t]
\centering
\includegraphics[width=\columnwidth]{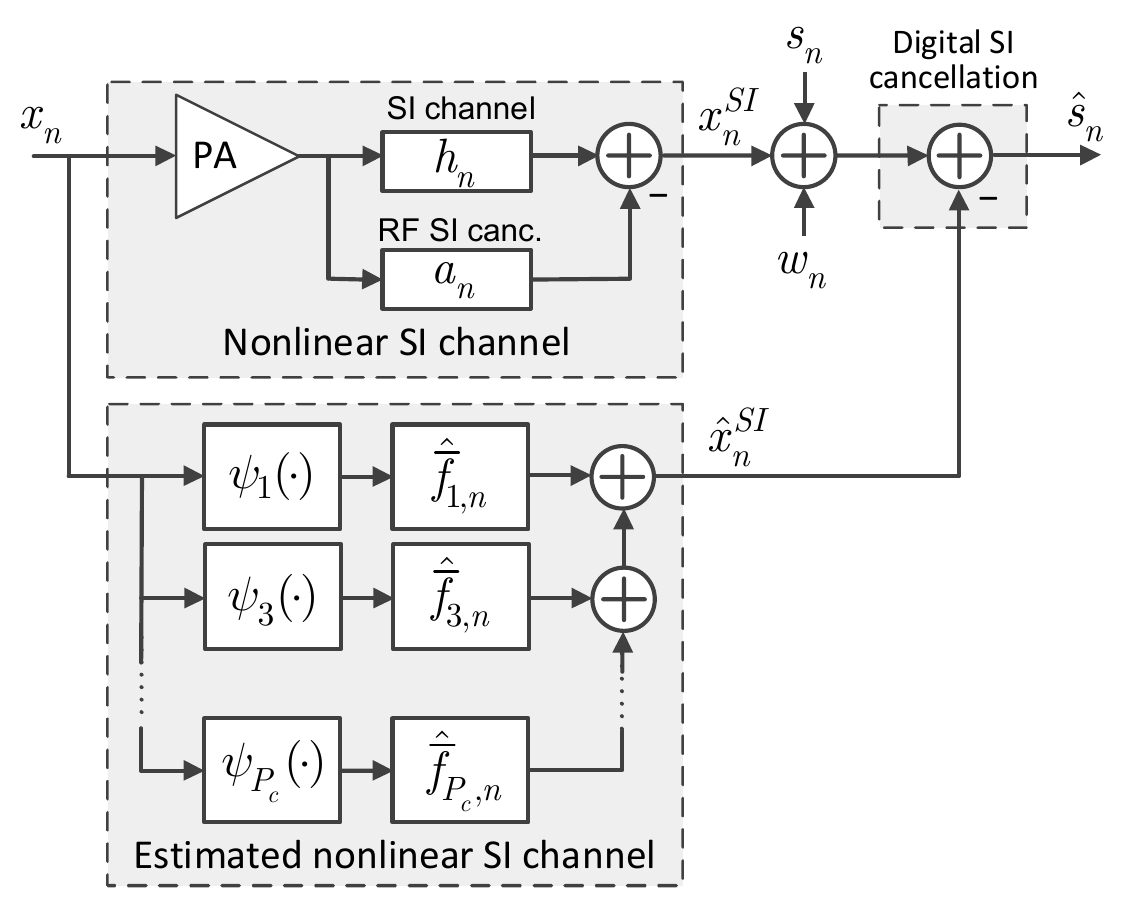}
\caption{A block diagram illustrating the transceiver modeling and the proposed canceller structure.}
\label{fig:diag_canceller}
\end{figure}

Despite the nonlinear nature of the model, it is linear in the parameters $\bar{f}_{{p},{k}}$, thus facilitating efficient estimation with, for example, linear least-squares methods. Notice also that the PH nonlinearity can model perfectly a variety of other PA-plus-channel models as well, for example the cascade of a polynomial nonlinearity and an LTI system, or the cascade of a Hammerstein nonlinearity and an LTI system, in addition to modeling the cascade of a PH nonlinearity and an LTI system. Therefore, choosing the PH nonlinearity as the cascade model is well justified in this sense also.

\subsection{Digital SI canceller parameter estimation}

The objective is now to estimate the parameters $\bar{f}_{{p},{k}}$ based on the above SI signal model, and then to regenerate the SI signal and subtract it from the overall received signal at digital baseband. A block diagram of the overall nonlinear SI channel, its modelling, regeneration, and digital cancellation is shown in Fig.~\ref{fig:diag_canceller}. There, the different branches, corresponding to nonlinear terms of different order, can be seen, alongside with the coefficients $\bar{f}_{{p},{n}}$. The total received signal before digital cancellation is now given as
\begin{align}
	x_n^\mathit{RF} = s_n+w_n+x_n^\mathit{SI} \text{,}
\end{align}
where $s_n$ is the actual received signal of interest, $w_n$ is the additive noise component, and $x_n^\mathit{SI}$ is the received self-interference signal whose model was given in \eqref{eq:x_n_SI}. Stemming from the previous modeling, the output of the digital SI canceller is given as
\begin{align}
	\hat{s}_n = x_n^\mathit{RF} - \hat{x}_n^\mathit{SI} \text{,}
\end{align}
with the \textit{self-interference estimate} $\hat{x}_n^\mathit{SI}$ given as
\begin{align}
	\hat{x}_n^\mathit{SI} = \sum_{\substack{{p} = 1 \\ {p \text{ } odd}}}^{{P_c}}
\sum_{{k} = -M_{c1}}^{{M}_{c2}} \hat{\bar{f}}_{{p},{k}} \psi_{p}(x_{{n}-{k}}) \text{.}
\end{align}
Here, $P_c$ is the nonlinearity order, and $M_{c1}$ and $M_{c2}$ are the non-causal and causal memory depths of the digital SI canceller.

To derive the estimator for $\bar{f}_{p,k}$, we first write vector representations of the relevant formulae with $L$ observed samples of $x_n^\mathit{RF}$:
\begin{gather*}
\mathbf{x}^\mathit{RF} = \mathbf{\Psi} \bar{\mathbf{f}}+\mathbf{s}+\mathbf{w} \text{, with} \nonumber\\
\mathbf{x}^\mathit{RF} \stackrel{\mathit{def}}{=} \begin{bmatrix} x_n^\mathit{RF} & x_{n+1}^\mathit{RF} \cdots x_{n+L-1}^\mathit{RF} \end{bmatrix}^\mathbf{T} \nonumber\\
\bar{\mathbf{f}} \stackrel{\mathit{def}}{=} \left[ \begin{smallmatrix} \bar{f}_{1,-M_{c1}} & \cdots & \bar{f}_{1,M_{c2}} & \bar{f}_{3,-M_{c1}} & \cdots & \bar{f}_{P_c,-M_{c1}} & \cdots & \bar{f}_{P_c,M_{c2}} \end{smallmatrix} \right]^\mathbf{T} \nonumber\\
\mathbf{\Psi} \stackrel{\mathit{def}}{=} \begin{bmatrix} \mathbf{\Psi}_1 & \mathbf{\Psi}_3 & \cdots  & \mathbf{\Psi}_{P_c} \end{bmatrix} \nonumber\\
\mathbf{\Psi}_p = \left[ \begin{smallmatrix}  \psi_p({n+M_{c1}}) & \psi_p({n+M_{c1}-1}) & \cdots & \psi_p({n-M_{c2}}) \\
\psi_p({n+M_{c1}+1}) & \psi_p({n+M_{c1}}) & \cdots & \psi_p({n-M_{c2}+1}) \\
\vdots & \vdots & \ddots & \vdots \\
\psi_p({n+M_{c1}+L-1}) & \psi_p({n+M_{c1}+L-2}) & \cdots & \psi_p({n-M_{c2}+L-1}) \\
\end{smallmatrix} \right] \nonumber
\end{gather*}

The least-squares estimator is then derived as the vector $\bar{\mathbf{f}}$ which minimizes the power of the digital SI canceller output $\hat{s}_n$, treating the received signal of interest $s_n$ as noise, as
\begin{align}
	\hat{\bar{\mathbf{f}}}^\mathit{LS} = \operatorname*{arg\,min}_\mathbf{f}  \left\| \mathbf{x}^\mathit{RF} - \hat{\mathbf{x}}^\mathit{SI} \right\|^2 = \operatorname*{arg\,min}_\mathbf{f} \left\| \mathbf{x}^\mathit{RF} - \mathbf{\Psi} \hat{\bar{\mathbf{f}}} \right\|^2 \text{.} \nonumber
\end{align}
Assuming full column rank in $\mathbf{\Psi}$, we obtain the well-known solution
\begin{align}
	\hat{\bar{\mathbf{f}}}^\mathit{LS} = (\mathbf{\Psi^H}\mathbf{\Psi})^{-1} \mathbf{\Psi^H} \mathbf{x}^\mathit{RF} \text{.} \nonumber
\end{align}

The basis matrix $\mathbf{\Psi}$ is known inside the device, so the estimation can be performed during actual data transmission, or by using embedded pilot signals within the data frame. In the latter case the calculation of the pseudoinverse $(\mathbf{\Psi^H}\mathbf{\Psi})^{-1} \mathbf{\Psi^H}$ can be performed a priori and stored, therefore reducing the real-time computational load. Adaptive estimation techniques, for example using the recursive LS (RLS) algorithm, usually avoid the matrix inversion, and are therefore practically more appealing compared to the above block LS solution when actual transmit data is used for estimation. These are straightforward to derive based on adaptive filter theory literature (e.g., \cite{Haykin96}) and are left to the reader.

\section{Numerical Results and Analysis}
\label{sec:results}
\vspace{-1mm}
In this Section, we perform full waveform simulations of the whole full-duplex transceiver with Matlab/Simulink to test and analyze the proposed nonlinearity model and its estimation and cancellation. The transceiver parameters are the same as in the system calculation example in Section~\ref{sec:challenges}. The simulated transmit and receive waveforms are OFDM signals with the parameters given in Table~\ref{table:simul_param}. The SI coupling channel between TX and RX antennas is modeled as a FIR filter, including the main path plus four multipath components. The power difference between the main component and the multipath components (K-factor) is approximately 36 dB \cite{Duarte12}. In the simulations, RF cancellation is implemented by subtracting the transmitted signal from the received signal with a small amplitude and phase mismatch such that the specified amount of total SI power reduction is obtained. For the PA, a Wiener model is used, which means that the model consists of a cascade of a 5th-order FIR filter modeling the memory effects and a polynomial modeling the actual nonlinear behaviour. Since there is a LTI channel at the input of the nonlinearity, the parallel Hammerstein model is not able to perfectly model the PA, and there will inherently be some residual SI due to a model mismatch.
%, and the maximum delay of the channel is 62.5 ns

In this paper, the used figure-of-merit is the SINR of the detector input signal. The PH nonlinear channel model has nonlinearity order 5, and the filter lengths of the PH model are 5 for all the branches. Furthermore, 10 OFDM symbols, or 3200 samples, are used for their estimation in each realization. For estimating the linear SI channel only, normal linear least-squares estimation is employed with the same number of samples and the same filter length.

\begin{table}[!t]
\renewcommand{\arraystretch}{1.3}
\caption{Additional parameters for the waveform simulator.}
\label{table:simul_param}
\centering
\begin{tabular}{|c||c|}
\hline
\textbf{Parameter} & \textbf{Value}\\
\hline
Constellation & 16-QAM\\
\hline
Number of subcarriers & 64\\
\hline
Number of data subcarriers & 48\\
\hline
Guard interval & 16 samples\\
\hline
Sample length & 15.625 ns\\
\hline
Symbol length & 4 $\mu$s\\ % Excluding GI!, With GI 5 \mu s
\hline
Signal bandwidth & 12.5 MHz\\
\hline
Oversampling factor & 4\\
\hline
PA memory length & 6\\
\hline
K-factor of the SI channel & 35.8 dB\\
\hline
SI channel length & 5\\
\hline
%Maximum delay of the SI channel & 62.5 ns\\
%\hline
\end{tabular}
\vspace{-4mm}
\end{table}

\begin{figure}[!t]
\centering
\includegraphics[width=\columnwidth]{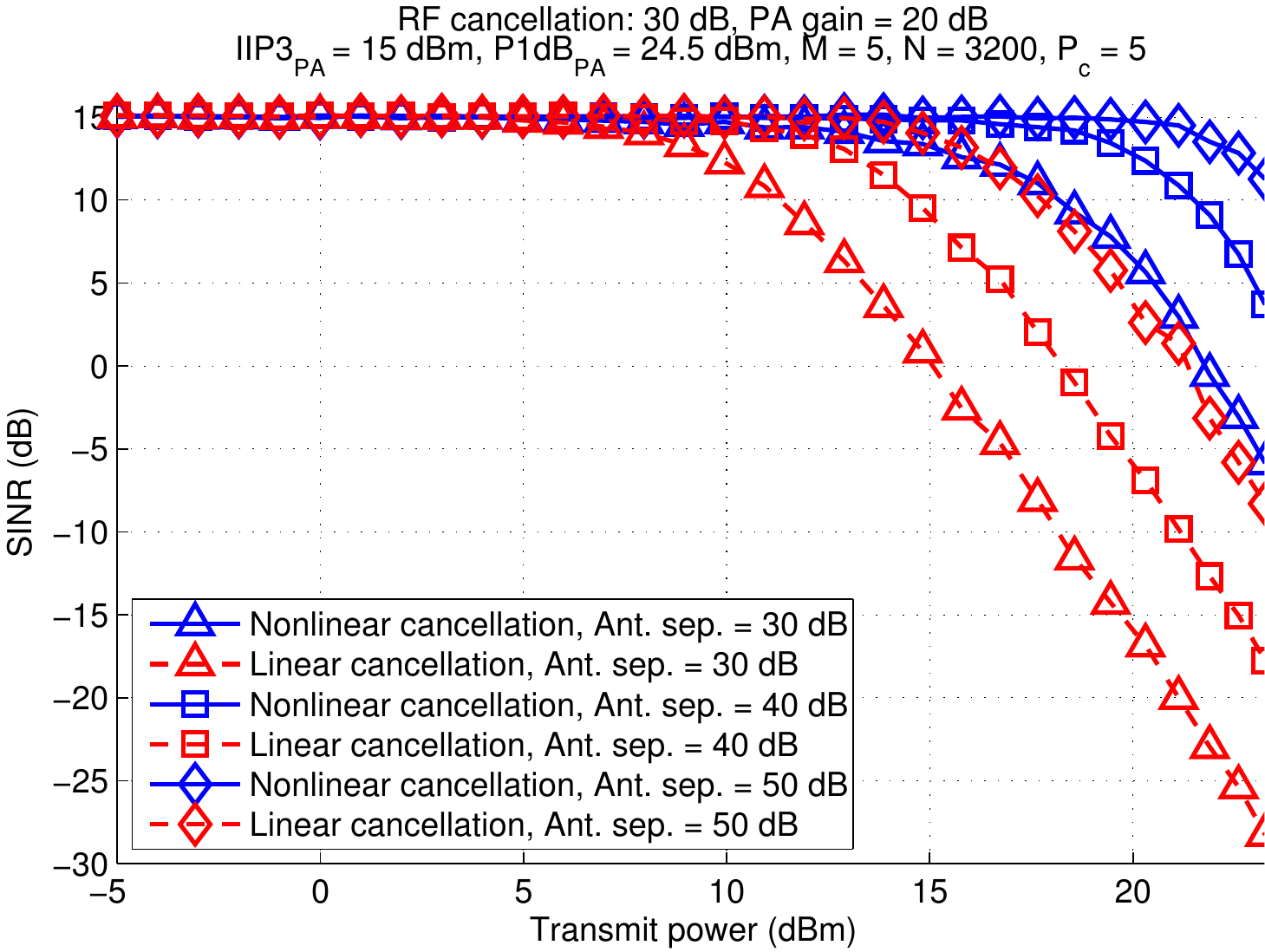}
\caption{SINR at detector input with and without nonlinear cancellation for various TX powers, and for different amounts of antenna separation.}
\label{fig:sinr_ptx}
\end{figure}

Figure~\ref{fig:sinr_ptx} shows the SINR vs. transmit power when only linear SI cancellation is employed, and when the proposed nonlinear SI cancellation structure is used. Both of these curves are plotted for three different values of antenna separation. The proposed nonlinear technique allows using approximately 10 dB higher transmit power than with linear SI cancellation, regardless of the amount of antenna separation. Furthermore, with antenna separation of 50 dB, the nonlinear SI cancellation technique achieves the same fidelity as without any SI up to a transmit power of 20 dBm. Thus, the proposed technique allows extending the operating range of the device in a significant manner. Furthermore, considering that the number of samples used for the estimation is only 3200, this is a promising result.% This kind of nonlinear self-interference cancellation methods have not been reported earlier in the full-duplex radio literature, thus emphasizing the novelty of this work.

\begin{figure}[!t]
\centering
\includegraphics[width=\columnwidth]{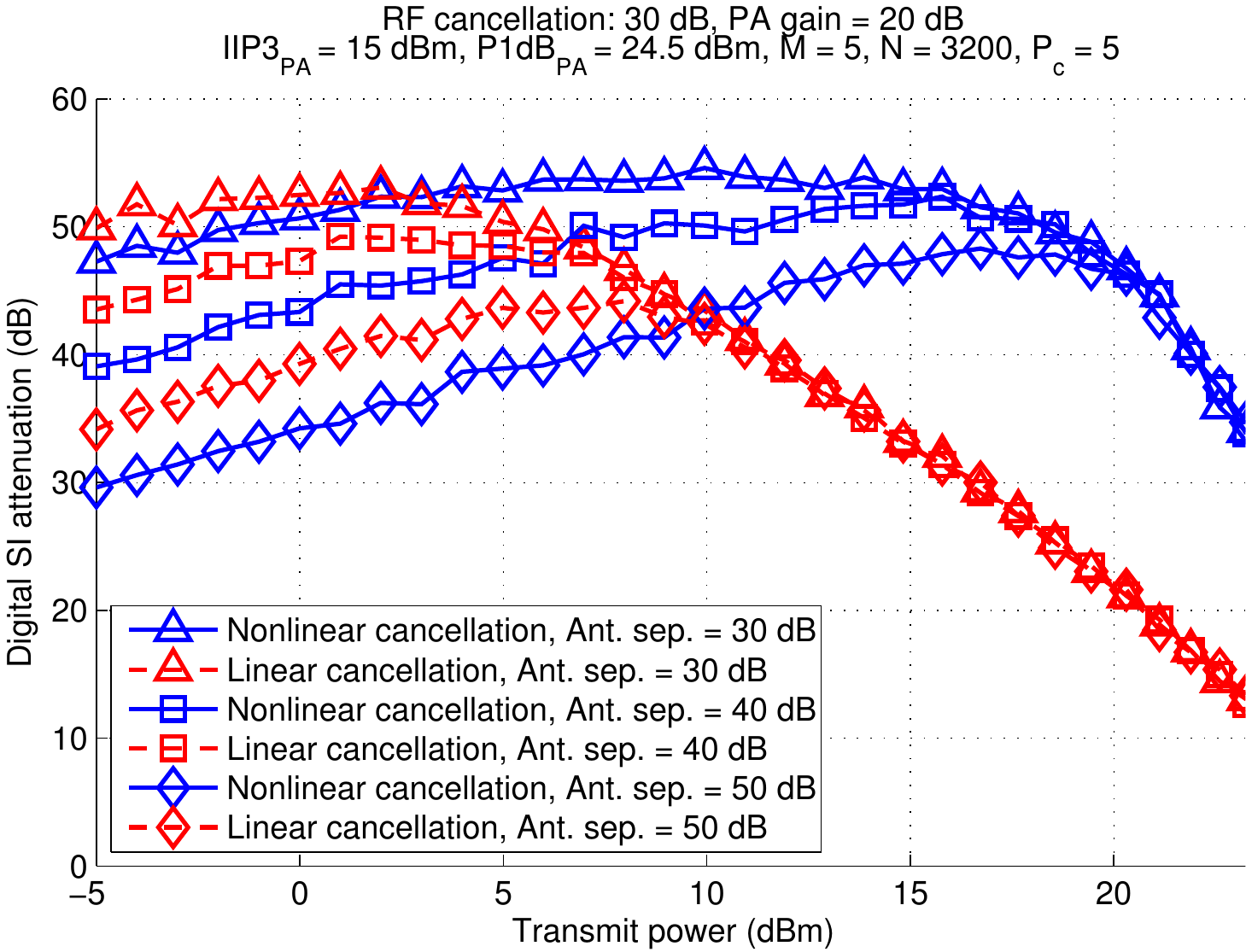}
\caption{The amount of achieved digital cancellation with and without nonlinear modeling for various TX powers, and for different amounts of antenna separation.}
\label{fig:dc_ptx}
\vspace{-5mm}
\end{figure}

The achieved amounts of digital SI attenuation, corresponding to Fig.~\ref{fig:sinr_ptx}, are shown in Fig.~\ref{fig:dc_ptx}. Here, the amount of achieved digital cancellation is defined as the decrease in the power of the total SI signal in the digital domain. It can be observed that in the linear operating region of the PA, a higher amount of digital cancellation is achieved with linear processing, as a lower number of parameters has to be estimated than with nonlinear processing, which suffers from over-parameterization. Taking a look back at Fig.~\ref{fig:p_all_example}, this behavior is natural since the power of the nonlinear distortion is well below the thermal noise floor with these lower transmit powers. However, as the transmit power increases, the power of the PA-induced nonlinear distortion becomes more significant, and the gain achieved by performing nonlinear SI cancellation increases. The point at which nonlinear SI cancellation becomes more beneficial than linear cancellation depends on the amount of analog SI attenuation. With very high transmit powers, the amount of achievable SI supression with the proposed nonlinear SI cancellation algorithm starts decreasing due to the quantization noise floor and model mismatch. The latter is caused by the difference between the considered Wiener PA model, and the parallel Hammerstein model used for estimating the nonlinearity coefficients.

\begin{figure}[!t]
\centering
\includegraphics[width=\columnwidth]{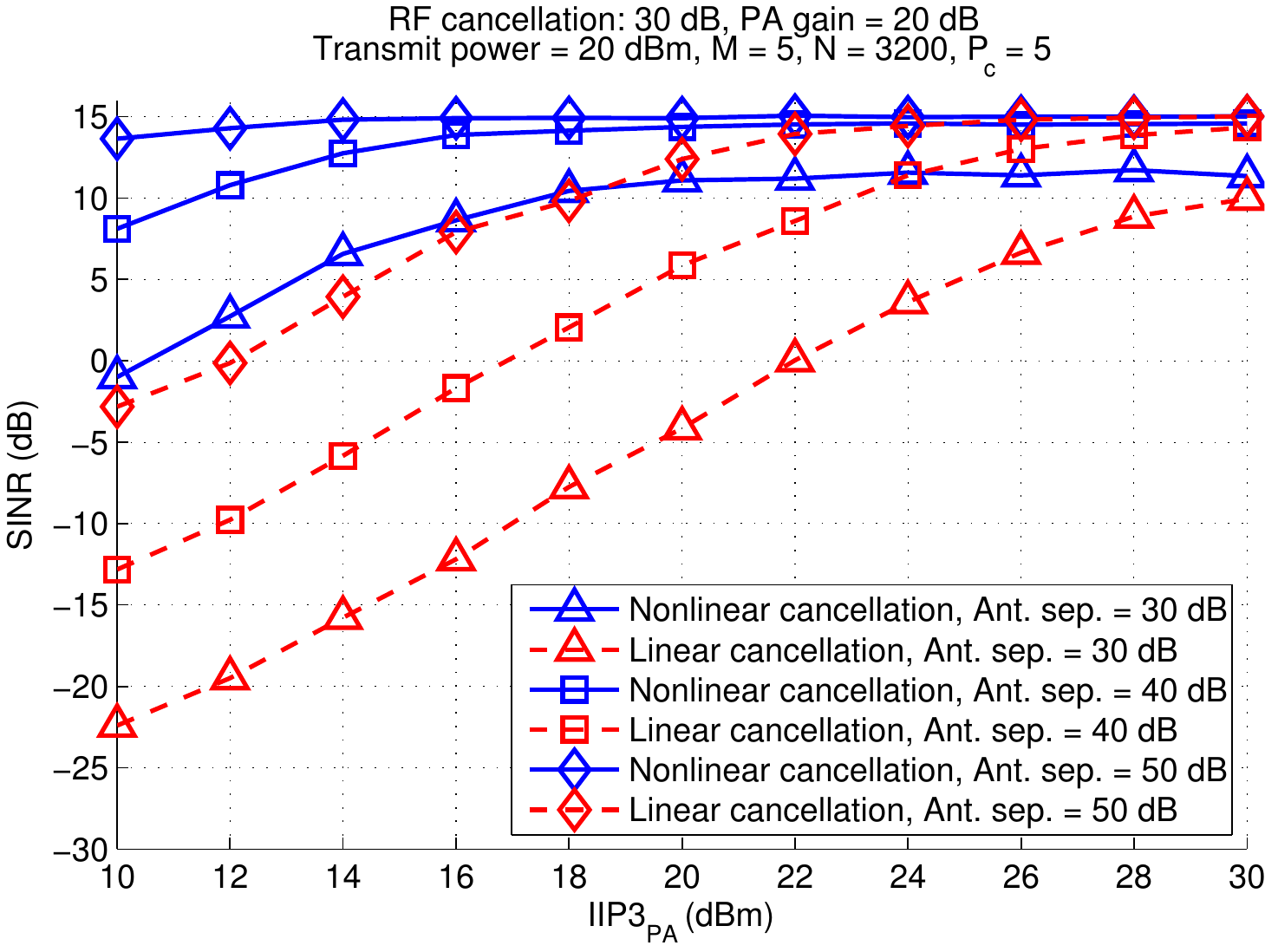}
\caption{SINR at detector input with and without nonlinear cancellation for various PA IIP3 figures.}
\label{fig:sinr_iip3}
\end{figure}

\begin{figure}[!t]
\centering
\includegraphics[width=\columnwidth]{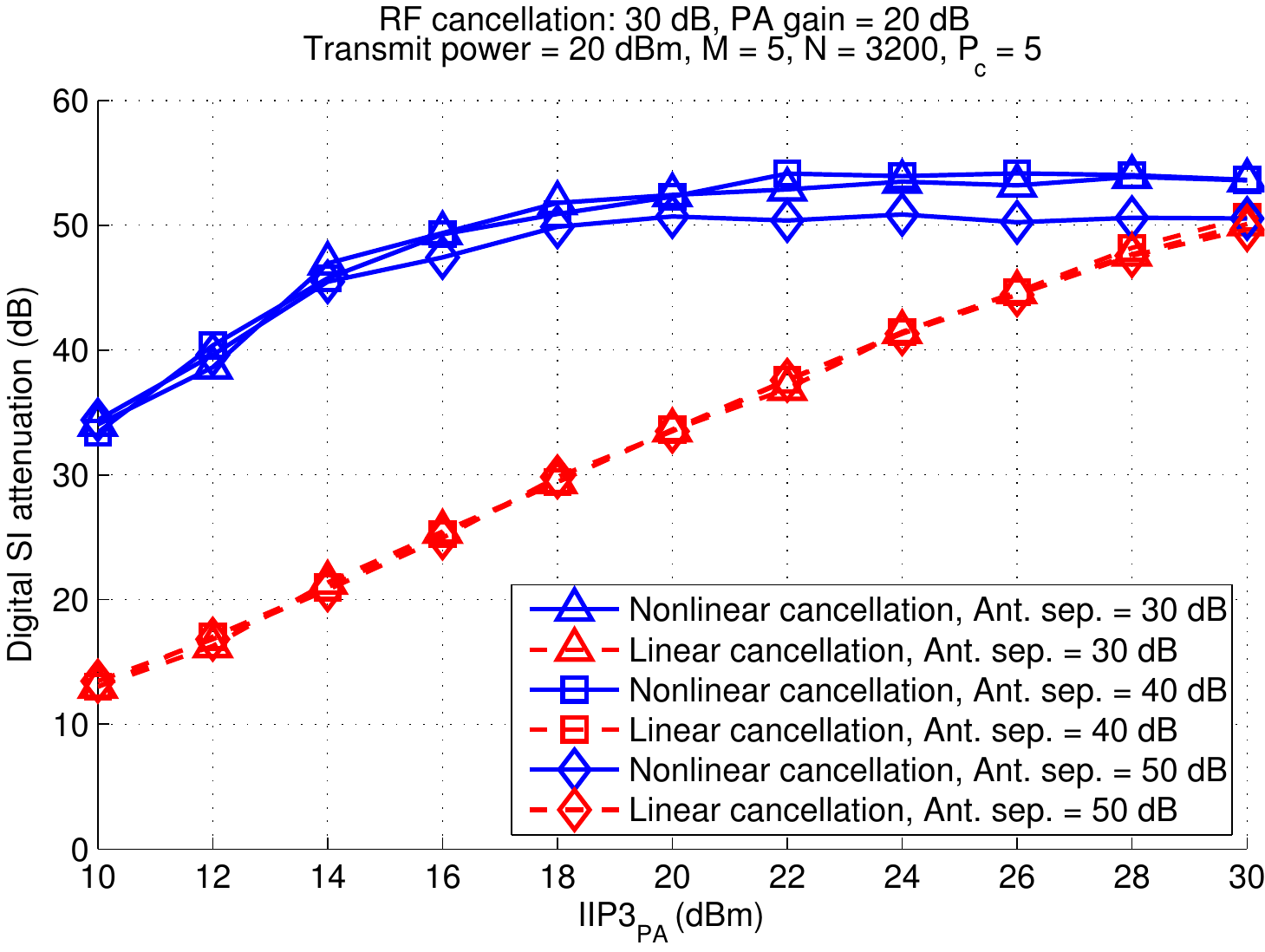}
\caption{The amount of achieved digital cancellation with and without nonlinear modeling for various PA IIP3 figures.}
\label{fig:dc_iip3}
\vspace{-3mm}
\end{figure}

Figure~\ref{fig:sinr_iip3} illustrates SINR vs. IIP3 of the PA, again with three different values of antenna separation. The purpose is to find out whether a lower-quality PA with a lower IIP3 figure could be used if nonlinear SI cancellation is employed. In these simulations, the value of the 1 dB compression point is always chosen as 9.5 dB higher than the IIP3 figure. Fig.~\ref{fig:sinr_iip3} confirms that a less linear PA can indeed be used with the proposed nonlinear SI cancellation algorithm, by demonstrating up to 15 dB reduction in IIP3 compared to using linear SI cancellation only. With higher values for antenna separation, it is possible to achieve the ideal SINR of 15 dB even with an IIP3 figure of 14 dBm when using the proposed nonlinear SI cancellation algorithm. With linear processing, this is not possible, as can be observed from Fig.~\ref{fig:sinr_iip3}. However, it is important to note that usually the linearity requirements for the PA are also set by the spectrum emission standards, and thus it might not be possible to decrease the IIP3 figure beyond a certain point.

The corresponding amount of achieved digital cancellation is shown in Figure~\ref{fig:dc_iip3}. It can be observed that the amount of antenna separation does not significantly affect the achievable digital SI attenuation with a transmit power of 20 dBm. Furthermore, with a less linear PA, the difference between the performances of linear and nonlinear cancellation algorithms is fairly large. However, as the linearity of the PA improves, the differences in the achieved digital SI attenuation decrease.

\section{Conclusion}
\label{sec:conc}
This article studied the effects and digital cancellation of PA induced nonlinear self-interference in full-duplex transceivers. It was first shown through transceiver system power calculations in a LTE uplink-like scenario that the PA nonlinearities are a significant problem with transmit powers exceeding about 10~dBm. A nonlinear digital self-interference cancellation technique was then proposed to handle both the linear and nonlinear self-interference simultaneously. Waveform simulations demonstrated the proposed canceller's ability to extend the usable transmit power range by at least 10~dB, or alternatively, to make it possible to use a lower-quality PA in the transmitter. This is seen as a major step ahead towards practical deployment of full-duplex radio communications with practical low-cost RF circuits, and in particular, with practical nonlinear power amplifiers. As future work, we will consider implementing an actual full-duplex transceiver with a typical nonlinear PA, and evaluate the performance of the proposed digital cancellation algorithm under realistic conditions.

% Can use something like this to put references on a page
% by themselves when using endfloat and the captionsoff option.
\ifCLASSOPTIONcaptionsoff
  \newpage
\fi

%\newpage

\bibliographystyle{./IEEEtran}
% argument is your BibTeX string definitions and bibliography database(s)
\bibliography{./IEEEabrv,./IEEEref}

% that's all folks
\end{document}